\documentclass[10pt, conference]{IEEEtran}

\IEEEoverridecommandlockouts

\usepackage{amsmath,amssymb,amsfonts}
\usepackage{textcomp}
\usepackage{xcolor}
\usepackage{graphicx,xcolor,array}
\usepackage{pgfplots}
\pgfplotsset{compat=1.14}
\usepackage{tikz}
\usetikzlibrary{arrows}
\usetikzlibrary{automata,positioning}
\usepackage{makecell}
\usepackage{upgreek}
\usepackage{graphicx}
\usepackage{subcaption}
\usepackage{mwe}

\usepackage{cite}      
\usepackage{float}
\usepackage{caption}

\usepackage{graphicx}
\usepackage{mwe}

\usepackage{psfrag}

\usepackage[utf8]{inputenc}
\usepackage{url}      
\usepackage{tikz}

\usepackage{amsmath}

\usepackage{amsmath}  
\usepackage[nodisplayskipstretch]{setspace}

\usepackage{amssymb}

\usepackage{algorithm}

\usepackage{algpseudocode}
\usepackage{pifont}

\newcommand{\shorten}[1]{}

\newcommand{\signed}
    {{\unskip\nobreak\hfill\penalty50
      \hskip2em\hbox{}\nobreak\hfil $\blacksquare$
      \parfillskip=0pt \finalhyphendemerits=0 \par}}

\begin{document}

\title{Slicing Scheduling for Supporting \\ Critical Traffic in Beyond 5G}

\author{\IEEEauthorblockN{Ali Esmaeily\IEEEauthorrefmark{1}, Katina Kralevska\IEEEauthorrefmark{1}, and Toktam Mahmoodi\IEEEauthorrefmark{2}}
\IEEEauthorblockA{
\IEEEauthorrefmark{1}Dept. of Information Security and Communication Technology \\ Norwegian University of Science and Technology (NTNU), Norway\\
Email: \{ali.esmaeily, katinak\}@ntnu.no
}
\IEEEauthorblockA{
\IEEEauthorrefmark{2}Dept. of Engineering, King’s College London, London, UK\\
Email: toktam.mahmoodi@kcl.ac.uk
}
}

\maketitle

\begin{abstract}
One of the most challenging services fifth-generation (5G) mobile network is designed to support, is the critical services in-need of very low latency, and/or high reliability. It is now clear that such critical services will also be at the core of beyond 5G (B5G) networks. While 5G radio design accommodates such supports by introducing more flexibility in timing, how efficiently those services could be scheduled over a shared network with other broadband services remains as a challenge. In this paper, we use network slicing as an enabler for network sharing and propose an optimization framework to schedule resources to critical services via puncturing technique with minimal impact on the regular broadband services. We then thoroughly examine the performance of the framework in terms of throughput and reliability through simulation. 
\end{abstract}
 {\bfseries {Keywords}}: B5G, eMBB, URLLC, coexistence, resource allocation, puncturing, critical traffic.

\IEEEpeerreviewmaketitle

\section{Introduction} \label{introduction}
The 5G mobile network came with the promise of ten times better performance in all directions~\cite{ERICSSON5G}. However, the main paradigm shift has been in supporting services from industry which otherwise have had a dedicated network. Supporting critical services was enabled with two main enablers: the possibility to offer ultra-high reliability and low-latency and the capability to share one network between services with different needs, known as network slicing. It is now clear that the industry support and critical services will be one of the main targets for \textit{beyond 5G} (B5G) networks.

The 5G standard has introduced classes of services in order to encapsulate different requirements. The enhanced Mobile Broadband (eMBB) was introduced as an advanced version of 4G mobile broadband with higher throughput, while the Ultra-Reliable Low-Latency Communication (URLLC) was introduced to capture the needs of critical industry data traffic. The goal of the eMBB service is to attain a high data rate while delivering an acceptable reliability level. In contrast, URLLC services require stringent latency and reliability constraints to support critical industries such as smart factories, autonomous driving, or remote surgery~\cite{mendis20195g}. Such supports are addressed in the 5G New Radio (5G-NR) standard, i.e., the 3rd Generation Partnership Project (3GPP) RAN1/RAN2~\cite{3gpp.38.331,3gpp.38.824}. These specifications entail methods for eMBB service to obtain a high data rate and at the same time introduce flexible numerology allowing shorter transmission time, which then could be used for an immediate transmission of a smaller amount of latency-sensitive data through grant-free access~\cite{3gpp5GNR}. Scheduling URLLC traffic over the same resources that were provisioned for eMBB introduces challenges for the eMBB traffic; hence the rich literature on the coexistence of eMBB and URLLC is reviewed in Section~\ref{sec:rw}. 

Network slicing is seen as one of the leading enabling technologies for sharing network resources among multiple tenants of a network (including vertical industries), which can provide services with diverse requirements in B5G. Accordingly, radio resource scheduling is crucial to efficiently utilize shared resources between slices in order to meet various services' requirements of the tenants \cite{esmaeily2021small}. Hence, 3GPP RAN1/RAN2 specifications facilitate the realization of slicing over 5G-NR via 1) RAN awareness feature to perform traffic administration for slices belonging to different tenants, and 2) policy enforcement and radio resource management for the RAN slices~\cite{Official_doc_NG.127}. Such incorporation between 3GPP RAN1/RAN2 and network slicing drives efficient, flexible, and controllable radio resource sharing among slices. 

In this paper, we use network slicing as a concept for sharing the network resources between URLLC and eMBB traffic. We consider using regular Transmission Time Intervals (TTIs) for eMBB traffic and short TTIs for URLLC traffic. While the eMBB traffic is scheduled and will be transmitted as planned, the URLLC traffic will be transmitted immediately by \textit{puncturing} the eMBB transmission slot. We extend an existing loss model in the literature~\cite{8486430} and accurately express it to capture the impact of this \textit{puncturing} on the eMBB throughput. To this end, the main contributions of this paper are as follows:
\begin{itemize}
    \item Characterizing the resource allocation problem for the coexistence of eMBB/URLLC traffic scheduling using the puncturing technique with the main objective of maximizing the minimum data rate of each eMBB user. 
    \item Precisely formulating the loss function definition to capture the impact of puncturing, resulting from overlapped URLLC traffic, on each eMBB user's throughput and per TTI and for every particular allocated radio resource to each eMBB user.
    \item Presenting an optimization framework ensuring the loss in eMBB throughput due to scheduling URLLC traffic is minimal; hence achievable data rate for the eMBB users is not affected significantly. We define a \textit{puncturing rate threshold} to limit such impact. Moreover, We benchmark our proposed solution with the state-of-the-art approaches. Simulation results confirm that the proposed method can 1) fulfill URLLC reliability requirements and 2) at the same time maintain the minimum achievable rate of the worst-case eMBB user, close to the minimum acceptable data rate for the eMBB users even for a high amount of incoming URLLC load. Worst-case eMBB user refers to the user located at the cell edge (with low allocated power or poor channel gain) or the most punctured user with the overlapped URLLC load. 
\end{itemize}

The remainder of this paper is organized as follows. Section~\ref{sec:rw} provides an overview of the literature on eMBB and URLLC coexistence. Section~\ref{sec:sm} presents the system model we use in this paper and problem formulation of the optimization framework. The simulation results are presented in Section~\ref{sec:sim}. Section~\ref{sec:con} concludes the paper.

\section{Related work} \label{sec:rw}
The conventional orthogonal-based radio resource allocation mechanism is not suited for the coexistence of URLLC and eMBB traffic~\cite{3gpp.21.915}. One of the proposals from the 3rd Generation Partnership Project (3GPP) to efficiently multiplex eMBB and URLLC data transmissions via the 5G-NR is the superposition/puncturing scheme. Superposition/puncturing~\cite{8476595} is performed by applying non orthogonal-based scheduling~\cite{7263349} of both eMBB and URLLC traffic on the same radio channel simultaneously.
Superposition/puncturing scheme is recognized as a promising technique to enable the coexistence of the eMBB and URLLC transmissions over the 5G-NR and thus has attracted much attention in academia and industry. Reference~\cite{8486430} models the impact of the URLLC transmission over the scheduled eMBB traffic via loss functions caused by the URLLC traffic. Reference~\cite{8638932} investigates the multiplexing of the eMBB and URLLC traffic in the Cloud RAN (C-RAN) environment. eMBB and URLLC traffic are transmitted via multicast and unicast transmission mode, respectively. The authors also provide a framework in order to maximize the revenue stream of the C-RAN provider. The study in~\cite{inproceedings} investigates mutual support of visual (over eMBB slice) and haptic (over URLLC slice) perceptions over cellular networks. Paper~\cite{8467353} suggests a two-sided matching game for a joint user association and resource allocation problem, using an analytic hierarchy process, which yields in enhancing resource allocation in the downlink eMBB and URLLC transmissions for a fog network.
The authors in~\cite{8497326} utilize a decomposition technique for the integrated eMBB and URLLC resource scheduling problem into two separate problems. For the case of eMBB, the authors employ the penalty successive upper bound minimization method over TTIs, and for the URLLC case, a transportation rule is applied over short TTIs. Paper~\cite{8647460} considers the efficiency of adopting the orthogonal-based and non orthogonal-based scheduling for the eMBB and URLLC traffic in a multi-cell C-RAN system. The work outcome reveals the advantage of using the orthogonal-based solution for degrading the mutual interference of the eMBB and URLLC traffic. The authors in~\cite{8476595} suggest a communication-theoretic basis for eMBB, mMTC, and URLLC services. The results showcase the performance of both orthogonal-based and non orthogonal-based slicing for different service types. The study in~\cite{10.1145/3297280.3297513} utilizes a matching game for the joint eMBB and URLLC traffic. The authors denote an optimization approach to maximize the minimum demanded eMBB data rate and, at the same time, analyze URLLC reliability constraints. Reference~\cite{8643428} presents a risk-sensitive strategy according to the conditional value at risk method for eMBB reliability and a chance constraint for URLLC reliability. The work in~\cite{6968269} provides an optimization problem obtained from an intelligent resource allocation scheme based on the puncturing approach by considering reliability for eMBB and URLLC services. The authors apply a deep reinforcement learning policy to discover the total number of punctured mini-slots of the whole eMBB users.

Unlike those works, which mainly focus on maximizing the sum rate of the eMBB users, this paper concentrates on maximizing individual minimum achievable data rate for the eMBB users. We describe the resource allocation problem for each eMBB user that experiences a negative impact on its data rate due to the incoming URLLC traffic. Such traffic punctures some or even all of the allocated resources to the eMBB user in a time slot.
\section{System model and problem formulation} \label{sec:sm}
\begin{figure*}[hbt!]
\centering
\includegraphics[scale=0.27]{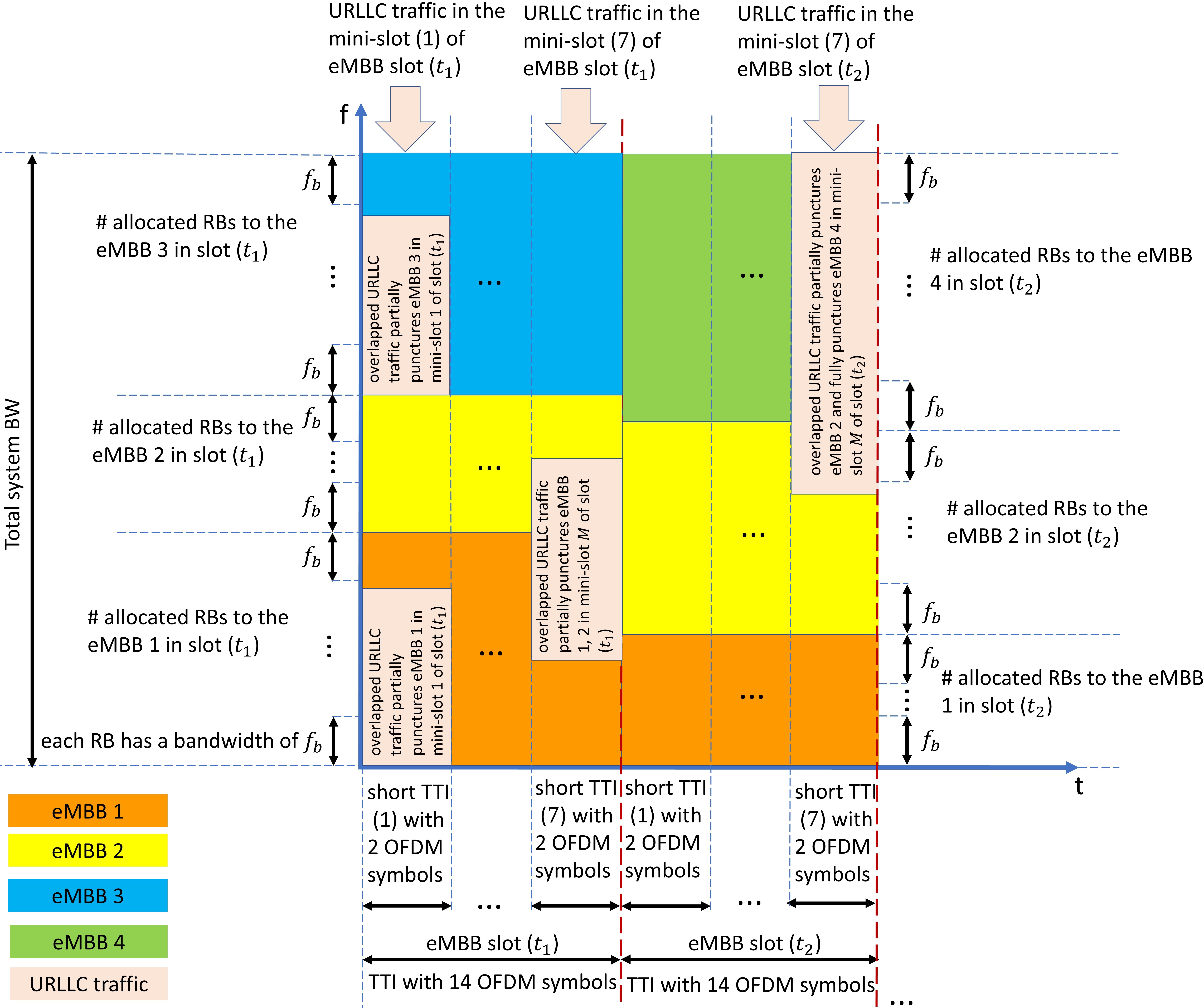}
\caption{eMBB/URLLC coexistence enabled by the puncturing mechanism for the numerology $\mu=0$.}
\label{fig:eMBB_URLLC_puncturing_approach}
\end{figure*}
\subsection{System Model}
In this network, we consider downlink eMBB and URLLC traffic, i.e., transmission from the network to the pieces of User Equipment (UEs) over a single gNB that can operate with single or multiple antennas $j \in \mathcal{J} = \{1, 2,...,J\}$. The gNB schedules the eMBB and URLLC traffic and transmits the corresponding data for each service type via its antennas towards eMBB and URLLC users over flat i.i.d Rayleigh fading channels. The gNB serves $k \in \mathcal{K} = \{1, 2,...,K\}$ number of eMBB and $n \in \mathcal{N} = \{1, 2,...,N\}$ number of URLLC UEs. The time domain is split into equally spaced time slots (TTIs) for the eMBB UEs' transmissions. Each time slot is subdivided into a fixed number of $M$ equally spaced mini-slots (short TTIs) where $m \in \mathcal{M} = \{1, 2,...,M\}$ denotes a mini-slot. In the frequency domain, the radio resources are divided into $b \in \mathcal{B} = \{1, 2,...,B\}$ Resource Blocks (RBs). Each RB $b$ contains 12 sub-carriers in the frequency domain and 14 OFDM symbols in the time domain. Since there is no strict latency requirement for serving the eMBB users, the RBs are allocated to them at the beginning of each time slot.  However, the sporadic URLLC requests can arrive at any time within a time slot, and due to the extreme latency requirement of such requests, the gNB needs to serve them immediately in a mini-slot instead of waiting for the next time slot. The gNB punctures previously scheduled eMBB transmissions in mini-slots by applying zero power to these transmissions to serve the URLLC requests promptly.

In 5G-NR, unlike 4G, the bandwidth of a RB, $f_{b}$, and time slot duration, TTI, are not fixed, and they are set according to specific values of sub-carrier spacing, $\Delta f$. Hence, there are several so-called numerologies in 5G-NR according to the values of $\Delta f$. Figure~\ref{fig:eMBB_URLLC_puncturing_approach} illustrates the puncturing mechanism for the coexistence of eMBB/URLLC traffic for the numerology zero-labeled as $\mu=0$ with $\Delta f$= 15 KHz, $f_{b}$= 180 KHz, TTI = 1 msec which contains 14 OFDM symbols, and each short TTI $\approx$ 142 $\mu$sec that occupies 2 OFDM symbols. Let consider the first mini-slot of the first time slot. The sporadic incoming URLLC traffic in the first mini-slot $m=1$ impacts the previously scheduled eMBB users with the allocated radio resources in this mini-slot. gNB decides to map the URLLC traffic to some of the eMBB UEs in this mini-slot. Hence, some of the resources of the eMBB UEs 1 and 3, $k=1, 3$ are punctured by the overlapped URLLC traffic.

Accordingly, the maximum achievable rate for an eMBB user $k$ at the time slot $t$ over the whole allocated RBs can be formulated as follows:
\begin{equation}\label{eq:data-rate-compact}
r_{k}^{eMBB}(t) = [\phi_{k}^{eMBB}(t) - \gamma_k^{eMBB}(t)] \times r_{k, peak}^{eMBB}(t)
\end{equation}
where the $\phi_{k}^{eMBB}(t)$ is the total amount of radio resources allocated to the eMBB user $k$ at time slot $t$, $\gamma_k^{eMBB}(t)$ is called the total loss function which indicates the fraction of punctured resources allocated to eMBB user $k$ at time slot $t$ due to the incoming URLLC requests, and $r_{k, peak}^{eMBB}(t)$ is the total achievable data rate of the eMBB user $k$ at time slot $t$. This formulation is general, and by following the Shannon channel capacity, it can be further extended to:
\begin{equation}\label{eq:data-rate-large}
\begin{split}
r_{k}^{eMBB}(t) = \sum\limits_{b=1}^B\Big[\Big(x_{kb}(t)f_{b} - \gamma_{kb}^{eMBB}(t)\Big)\times\\log_{2}\Big(1 + \frac
{\sum\limits_{j=1}^J p_{kb}^{j}(t)h_{kb}^{j}(t)}{\sigma^2}\Big)\Big]
\end{split}
\end{equation}
where $x_{kb}(t)$ is the resource allocation coefficient, $x_{kb}(t)$ = 1 denotes that the RB $b$ is allocated to the eMBB user $k$ at time slot $t$ and $x_{kb}(t)$ = 0 shows no allocation; $f_{b}$ is the bandwidth of the RB $b$; $p_{kb}^{j}(t)$ is the transmission power from the antenna $j$ of the gNB over the RB $b$ to the eMBB user $k$ at time slot $t$; $h_{kb}^{j}(t)$ is the Rayleigh fading channel gain of the transmission from the antenna $j$ of the gNB over the RB $b$ to the eMBB user $k$ at time slot $t$; $\sigma^2$ is the noise power; and finally $\gamma_{kb}^{eMBB}(t)$ indicates the fraction of punctured RB $b$ that is allocated to eMBB user $k$ at time slot $t$. Now, let $D_m(t)$ be a random variable indicating the number of incoming URLLC packets in the mini-slot $m$ of time slot $t$. Hence, the total incoming URLLC packets in the time slot $t$ is equal to $D(t) = \sum\limits_{m=1}^M D_m(t)$. As a result, the $\gamma_k^{eMBB}(t)$ can be formulated as follows: 
\begin{equation}\label{eq:loss-function}
\begin{split}
\gamma_k^{eMBB}(t) = \sum\limits_{b=1}^B \gamma_{kb}^{eMBB}(t) \\= \Big[\sum\limits_{b=1}^B x_{kb}(t)f_{b} \times \rho_{kb}(t)\frac
{ D(t)}{|B| \times |M|}\Big]
\end{split}
\end{equation}
where $\rho_{kb}(t) \in \mathcal[0, 1]$ indicates the weight of puncturing; and $|B| \times |M|$ presents the total system capacity in terms of frequency-time resources. The URLLC traffic is upper bounded by total system capacity, i.e., $D(t) \le |B| \times |M|$. The $\rho_{kb}(t)$ identifies the pattern of overlapping total URLLC traffic in the time slot $t$ on the eMBB user $k$ resources in order to utilize (puncture) them for the URLLC transmission. According to the pattern of puncturing the eMBB resources, the $\gamma_k^{eMBB}(t)$ function can be approximated as a regular algebraic function. In this paper, we define $\gamma_k^{eMBB}(t)$ as first and second-degree non-decreasing polynomial known as linear and convex quadratic functions, respectively, where $\gamma_k^{eMBB}(t) \in \Big[0, \sum\limits_{b=1}^B x_{kb}(t)f_{b}\Big]$. Hence, for each eMBB user $k$ in time slot $t$ if: 
\begin{itemize}
    \item $\gamma_k^{eMBB}(t)=0$, no puncturing occurs;
    \item $0 < \gamma_k^{eMBB}(t) < \sum\limits_{b=1}^B x_{kb}(t)f_{b}$, partial puncturing happens;
    \item $\gamma_k^{eMBB}(t)=\sum\limits_{b=1}^B x_{kb}(t)f_{b}$, full puncturing appears.
\end{itemize}
It should be noted that the individual achievable data rate for the eMBB user $k$  in time slot $t$ holds a higher value if this user suffers from a resource deduction scheme following the convex function rather than the linear function.

Until now, we have only considered the latency requirement for the incoming URLLC requests by scheduling them on top of eMBB transmissions. Regarding the reliability requirement of URLLC traffic, let $\theta_{max}$ be the outage probability threshold and $\eta$ be the URLLC packet size, then the reliability of URLLC UEs can be represented as~\cite{8643428}:
\begin{equation}\label{eq:error-probability}
\begin{split}
Pr(error) = Pr \Big\{\sum\limits_{n=1}^N\sum\limits_{k=1}^K\Big[\frac{\gamma_{k}^{eMBB}(t)}{f_{b}N}\times\\log_{2}\Big(1 + \frac
{\sum\limits_{j=1}^J p_{nb}^{j}(t)h_{nb}^{j}(t)}{\sigma^2}\Big) \Big]\le \eta D(t)\Big\} \le \theta_{max}.
\end{split}
\end{equation}
Under the joint eMBB/URLLC resource allocation problem, the objective is to maximize the data rate for each of the eMBB UEs and at the same time fulfill the URLLC UEs' requirements in terms of extra low delay and high reliability:
\begin{subequations}
\begin{equation}\label{eq:data-rate-optimization}
\max_{
p, \gamma} \min_{k \in \mathcal{K}} \mathbb{E} \{\sum\limits_{t=0}^T r_{k}^{eMBB}(t)\} 
\end{equation} 
\begin{equation}\label{eq:data-rate-optimization-condition-one}
subject \, to \quad Pr(error) \le \theta_{max}
\end{equation}
\begin{equation}\label{q:data-rate-optimization-condition-two}
 \sum\limits_{k=1}^K \sum\limits_{b=1}^B \sum\limits_{j=1}^J p_{kb}^{j}(t) \le P_{max} 
\end{equation}
\end{subequations}
where the $P_{max}$ is the maximum transmission power from the gNB towards all types of the UEs.
\subsection{Solving the coexistence optimization problem}
Here we present the proposed algorithm to find an optimal solution for Eq. (\ref{eq:data-rate-optimization}). In this algorithm, first, we set the minimum acceptable data rate $R_{min}$ for the eMBB users. Subsequently, in each time slot $t$ we define a \textit{puncturing rate threshold} $th^{eMBB}(t)$ according to the loss functions for all eMBB users. The selection criteria for calculating $th^{eMBB}(t)$ is as follows:
\begin{equation}\label{eq:puncturing-rate-threshold}
 th^{eMBB}(t)= \begin{cases} \max_{\forall k \in \mathcal{K}} \{\ \gamma_k^{eMBB}(t)\},&\\ \hspace{12mm} 0 \le \gamma_k^{eMBB}(t)< \sum\limits_{b=1}^B x_{kb}(t)f_{b};\\ 
\max_{\forall k \in \mathcal{K}} \{\ \gamma_k^{eMBB}(t)\}-\textit{offset},& \\ \hspace{12mm} \gamma_k^{eMBB}(t) = \sum\limits_{b=1}^B x_{kb}(t)f_{b};\end{cases}
\end{equation}
where \textit{offset} indicates a constant value to tune $th^{eMBB}(t)$ if the second condition holds. It is worth noting that the first condition for defining $th^{eMBB}(t)$ is much more likely to happen than the second one. After setting a value for $th^{eMBB}(t)$, we proceed to calculate the achievable rate for each eMBB user $k$ in the time slot $t$. Next, we check whether the achievable rate is less than $R_{min}$. If $r_{k}^{eMBB}(t) < R_{min}$, then we map the incoming URLLC load to another possible eMBB user $k'$ with the allocated RB $b'$ if at least one of the following conditions is fulfilled:
\begin{itemize}
    \item higher power value,  i.e. $p_{k'b'}^{j'}(t) > p_{kb}^{j}(t)$; 
    \item larger channel gain value, i.e. $h_{k'b'}^{j'}(t) > h_{kb}^{j}(t)$;
    \item lower loss function, i.e. $\gamma_{k'}
^{eMBB}(t) < \gamma_k^{eMBB}(t)$; 
\end{itemize}
otherwise we hold with the current eMBB user $k$. In other words, the challenge corresponds to the minimum rate belonging to the most punctured eMBB users, which negatively impacts the performance of the system if the minimum rate would be less than the $R_{min}$. Hence, tracking each eMBB user rate is crucial in each time slot within a frame and for the whole transmission period. As a result, the optimization algorithm protects those eMBB users with low power allocation, bad channel quality, and less allocated RBs to avoid worsening their date rate by over-puncturing. Algorithm \ref{alg:data-rate-optimization-algorithm} summaries the steps.
\begin{algorithm}
\caption{Algorithm for eMBB/URLLC coexistence}
\label{alg:data-rate-optimization-algorithm}
\begin{algorithmic}[1]
    \State \textbf{Input:}{\, $t \in T, b\in \mathcal{B}, k \in \mathcal{K}, j \in \mathcal{J}, p_{kb}^{j}(t), h_{kb}^{j}(t), \gamma_{k}^{eMBB}(t)$}
    \State \textbf{Output:}{\, Solution to Eq. (\ref{eq:data-rate-optimization}) for  eMBB/URLLC coexistence}\\
    Set $R_{min}$\\
    Define $th^{eMBB}(t)$ according to Eq. (\ref{eq:puncturing-rate-threshold})
    \For {$t \in T$}
        \For{$k \in K$}
        \For{$j \in J$}
        \State Calculate $r_{k}^{eMBB}(t)$ based on $th^{eMBB}(t)$ 
        \If{$r_{k}^{eMBB}(t) < R_{min}$}
        \State Map the URLLC load to eMBB user $k'$ in case $p_{k'b'}^{j'}(t) > p_{kb}^{j}(t)$,  $h_{k'b'}^{j'}(t) > h_{kb}^{j}(t)$, or $\gamma_{k'}^{eMBB}(t) < \gamma_k^{eMBB}(t)$
        \Else
        \State Puncture the current eMBB user $k$
        \EndIf
        \EndFor
        \EndFor
    \EndFor
\end{algorithmic}
\end{algorithm}
\begin{table}[hbt!]
\centering
\caption{Simulation parameters.}
\begin{tabular}[b]{| c | c |}
  \hline
  \textbf{\scriptsize Simulation parameter} & \textbf{\scriptsize Value}\\ 
  \hline
    \scriptsize Cell radius(m) & \scriptsize 500 \\ 
  \hline
  \scriptsize Number of mini-slots & \scriptsize 7 \\ 
  \hline
   \scriptsize Number of OFDM symbols per mini-slot & \scriptsize 2 \\ 
  \hline
 \scriptsize Number of eMBB users & \scriptsize 5 \\ 
  \hline
 \scriptsize URLLC traffic model & \scriptsize Poisson process\\ 
  \hline
 \scriptsize $f_{b}$ (KHz) & \scriptsize 180 \\ 
  \hline
 \scriptsize Total BW (MHz)  & \scriptsize 20 \\ 
  \hline
\scriptsize Min guard band for numerology $\mu =0$ (KHz)& \scriptsize 692.5\\ 
  \hline		
\scriptsize Number of resource blocks & \scriptsize 103\\ 
  \hline  
\scriptsize $R_{min}$ (Mbps)& \scriptsize 5\\ 
  \hline
\scriptsize $P_{max}$ (dBm)& \scriptsize 40\\ 
  \hline
\scriptsize Time slot length (msec)& \scriptsize 1\\ 
  \hline
  \scriptsize Mini-slot length ($\mu$sec)& \scriptsize 142 \\ 
  \hline
\scriptsize Time frame length (msec)& \scriptsize 10\\ 
  \hline
\scriptsize URLLC packet size (Bytes)& \scriptsize 50\\ 
  \hline
\end{tabular}
\label{table:simulation_parameter}
\end{table}
\section{Performance evaluation} \label{sec:sim}
In this section, we verify the efficiency of our proposed algorithm through simulations and evaluate the performance. Our objective is to show the increase of the individual minimum achievable data rate for each eMBB user in the following analysis. We analyze and simulate the RAN domain using MATLAB R2019b with the CVX toolbox. In our simulated RAN, we consider one gNB located at the center of the cell coverage zone with a 500 m radius. The gNB operates on 20 MHz in the downlink mode, which serves several eMBB and URLLC UEs that are randomly distributed within the coverage zone. Besides, the gNB schedules eMBB and URLLC traffic in the downlink transmission over flat i.i.d Rayleigh fading channels. Table~\ref{table:simulation_parameter} summarizes the main simulation parameters. We benchmark the performance of our proposed solution with the well-known state-of-the-art approaches, including: 1) Punctured Scheduling (PS)~\cite{8287951}: PS selects the RBs with the highest MCS allocated to eMBB users and punctures them in order to serve URLLC traffic; 2) Random Scheduler (RS)~\cite{8486430}: RS serves the incoming URLLC traffic by randomly selecting pre-allocated RBs to the eMBB users; and 3) Equally Distributed Scheduler (EDS)~\cite{10.1145/3297280.3297513}: EDS serves the incoming URLLC traffic by equally choosing pre-allocated RBs to each of the eMBB users.
\begin{figure*}[hbt!]
        \centering
        \begin{subfigure}[b]{0.475\textwidth}
            \centering
            \includegraphics[width=\textwidth]{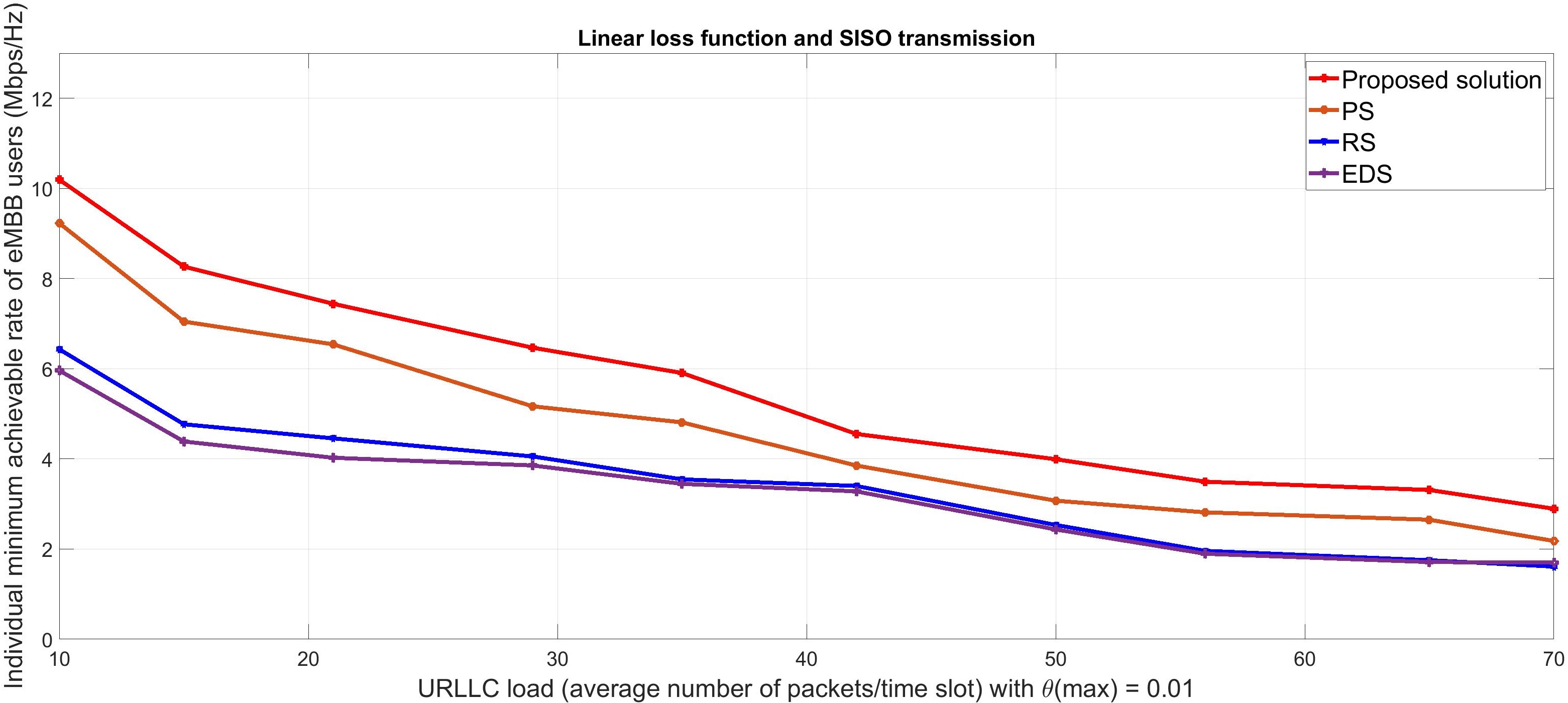}
            \caption[Network2]%
            {{\small \textit{(linear, SISO) regime}.}}    
            \label{fig:linear-siso}
        \end{subfigure}
        \hfill
        \begin{subfigure}[b]{0.475\textwidth}  
            \centering 
            \includegraphics[width=\textwidth]{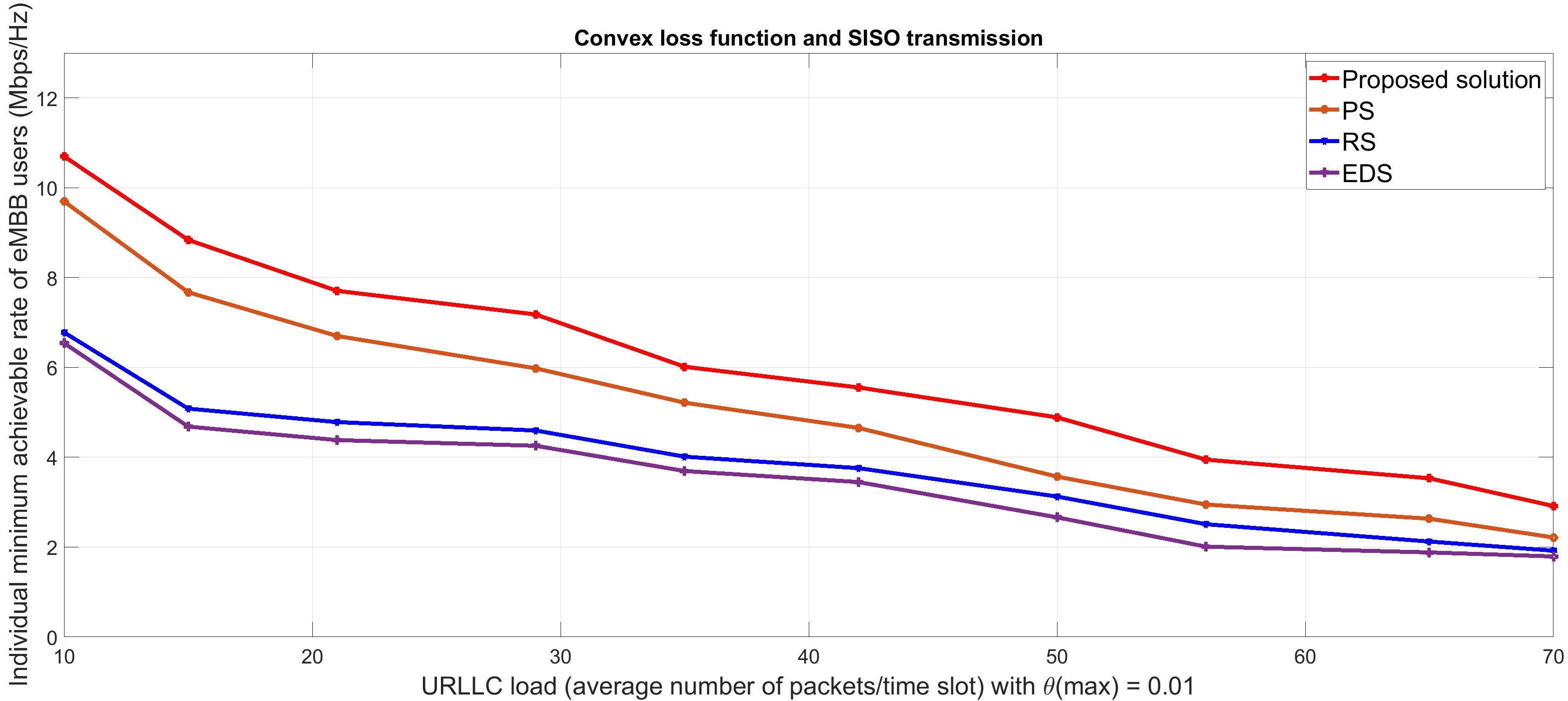}
            \caption[]%
            {{\small \textit{(convex, SISO) regime}.}}    
            \label{fig:convex-siso}
        \end{subfigure}
        \vskip\baselineskip
        \begin{subfigure}[b]{0.475\textwidth}   
            \centering 
            \includegraphics[width=\textwidth]{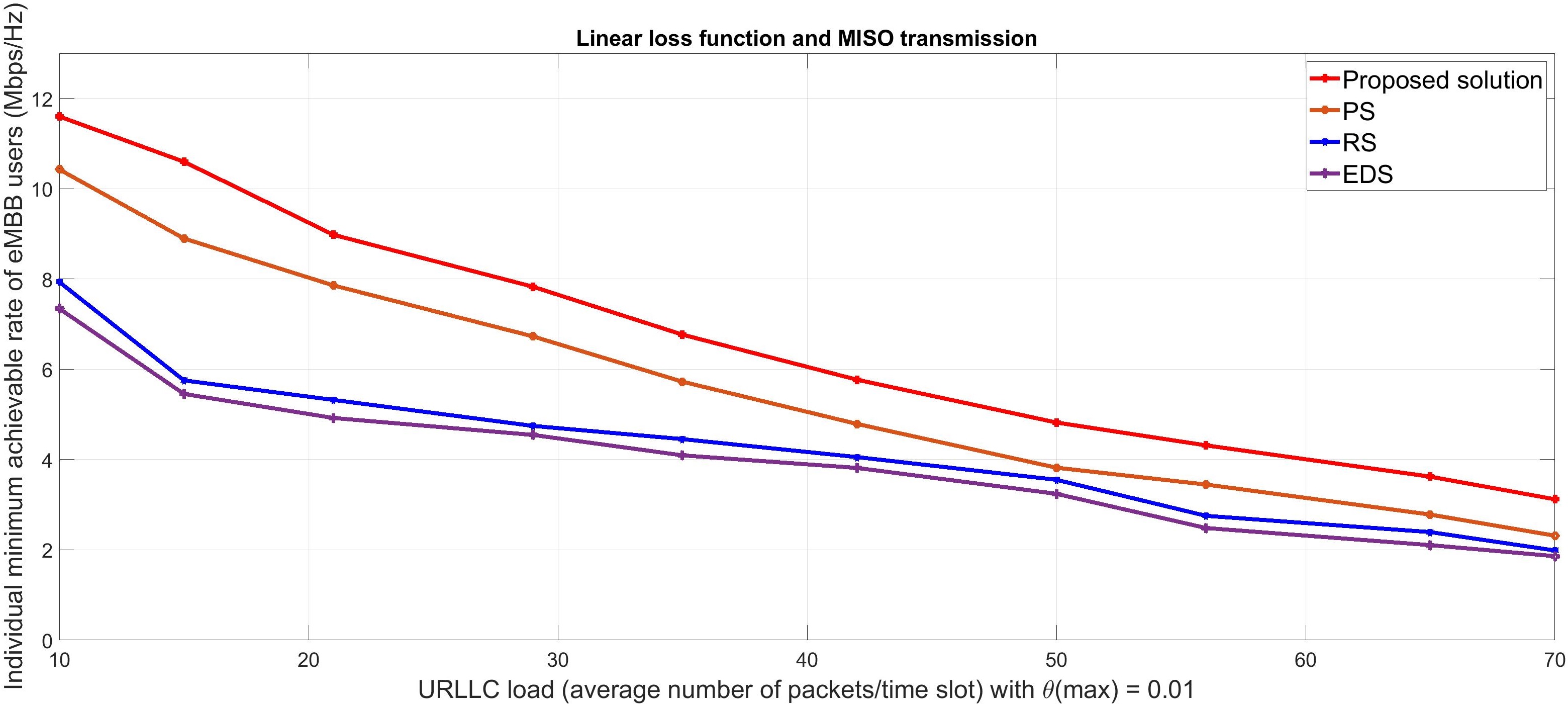}
            \caption[]%
            {{\small \textit{ (linear, MISO) regime}.}}  
             \label{fig:linear-miso}
        \end{subfigure}
        \hfill
        \begin{subfigure}[b]{0.475\textwidth}   
            \centering 
            \includegraphics[width=\textwidth]{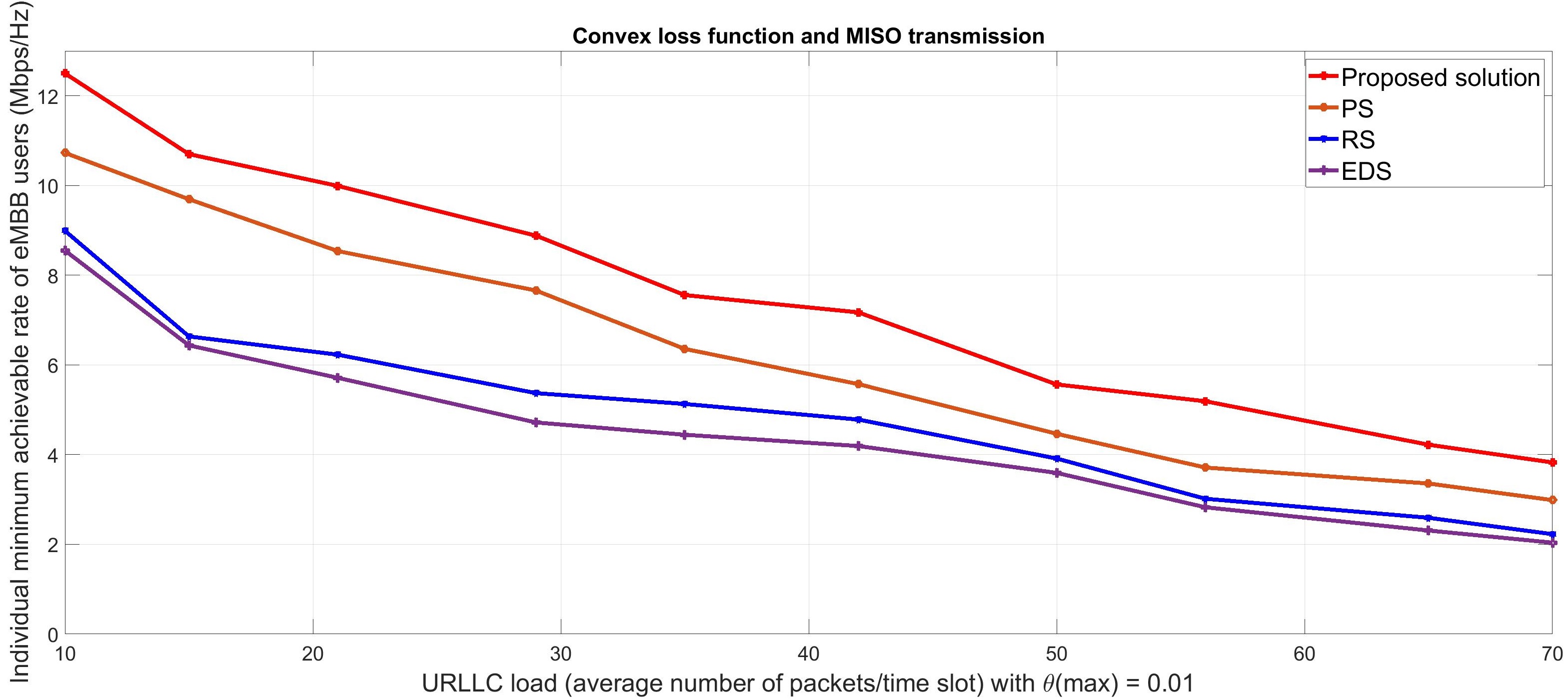}
            \caption[]%
            {{\small \textit{(convex, MISO) regime}.}}    
            \label{fig:convex-miso}
        \end{subfigure}
        \caption[]
        {\small Individual minimum achievable eMBB data rate for different number of URLLC packets per time slot.}  
        \label{fig:embb-data-rate}
    \end{figure*}
\subsection{eMBB data rate influenced by puncturing with URLLC load}
We first investigate the performance of the proposed algorithm in terms of resource allocation for the individual minimum achievable rate of the eMBB users. The deduction of the allocated resources to the eMBB users is represented by the $\gamma_k^{eMBB}(t)$ function in each time slot. We assume that the type of this function for the simulation environment is either a second-degree non-decreasing polynomial linear or convex quadratic function. Moreover, gNB can operate either with single or multiple numbers of antennas towards eMBB and URLLC users. We also consider that both eMBB and URLLC users are equipped with only a single antenna for data transmission. Hence the operation in downlink between the gNB and the users happens either in Single- or Multiple- Input Single Output configurations known as SISO and MISO, respectively. Figure~\ref{fig:embb-data-rate} illustrates four different \textit{regimes} that may happen via transmission of the data in the downlink in the form of \textit{(type of $\gamma_k^{eMBB}(t)$, type of the transmission configuration)}. For each \textit{regime}, we study the individual minimum achievable eMBB data rate per user via the proposed optimization algorithm, PS, RS, and EDS solutions. By increasing the URLLC load per time slot, depending on the scheduling strategy, some or all of the eMBB users may be influenced by puncturing. Particularly, in the \textit{(linear, SISO) regime} illustrated in Figure~\ref{fig:linear-siso}, for a number of 40 URLLC packets per time slot (considered as a mid-range number of URLLC packets per time slot), the minimum achievable rate for each eMBB user can reach up to 3.3, 3.5, 4.2, and 5.1 Mbps for the EDS, RS, PS, and our proposed solution respectively. By applying the optimization algorithm, gNB searches for at least one possible pre-scheduled eMBB candidate with higher allocated power, higher channel gain, or lower loss function to map full or partial URLLC load to that eMBB user while at the same time satisfying the minimum acceptable data rate for the eMBB users. The optimization process enhances even the worst-case eMBB user data rate to achieve up to 5.1 Mbps which is still greater than the $R_{min}$. The performance of the proposed algorithm is also prominent by increasing the minimum data rate up to 10.1 Mbps for the low amount of URLLC packets per time slot (10 packets). The same logic follows for the other \textit{regimes} as well. The most reliable case is \textit{(convex, MISO) regime}, presented in Figure~\ref{fig:convex-miso}. This \textit{regime} holds the convex loss function with less puncturing impact on the eMBB users than the linear loss function, and gNB operates with multiple antennas towards all users in the downlink. In the \textit{(convex, MISO) regime}, the proposed algorithm can improve the minimum data rate for the worst-case eMBB user up to 12.5 Mbps for 10 URLLC packets per time slot while PS, RS, and EDS can ultimately achieve up to 10.8, 9, and 8.8 Mbps respectively for the same user. It is worth considering that the efficiency of the proposed algorithm is noticeable even for a high amount of URLLC load with the rate of up to 57 packets per time slot, where the individual minimum achievable data rate is equal to the $R_{min}$. Due to the sporadic nature of such packets, the probability of having a very high number of URLLC packets per time slot is low, and thus the proposed algorithm is close to real scenarios. Besides, we assume the size of a URLLC packet is 50 bytes; however, smaller packet sizes are also expected, which results in less puncturing of the eMBB users. By keeping the individual minimum data rate close to the $R_{min}$, the network guarantees that each eMBB user receives at least minimum resources, which are required for normal web browsing and light video streaming. However, full HD video streaming with very high resolution demands some buffer time. In fact, with this strategy, the network does not allow to fully puncture eMBB users, and it keeps the data rate at a minimum level to avoid reducing the individual eMBB data rate significantly. The proposed algorithm outperforms PS, RS, and EDS solutions in different regimes under the same amount of URLLC load per time slot.

\subsection{eMBB reliability region for different URLLC load}
\begin{figure}[hbt!]
\centering
\includegraphics[scale=0.145]{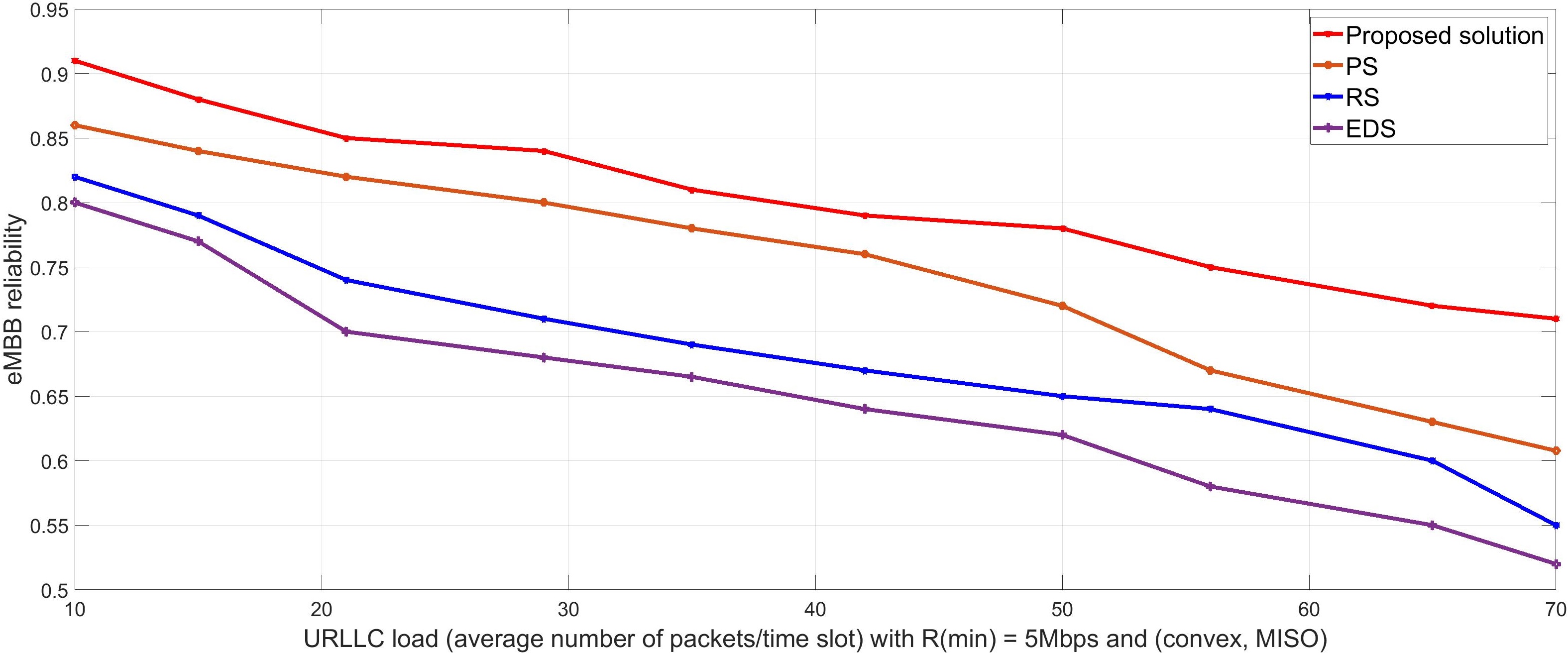}
\caption{eMBB reliability for different number of URLLC packets per slot with $R_{min}$ = 5 Mbps and \textit{(convex, MISO) regime}.}
\label{fig:reliability}
\end{figure}

Here we analyze the reliability of the eMBB users. We set $R_{min}$ equal to 5 Mbps and consider the most reliable transmission \textit{(convex, MISO) regime}. As Figure~\ref{fig:reliability} illustrates, applying the proposed algorithm during the transmission towards the eMBB users delivers a more reliable communication compared to the other scheduling policies in the downlink. Specifically, the eMBB users experience 91\% reliable transmission for 10 incoming URLLC packets per slot while PS, RS, and EDS can provide reliable transmission up to 86\%, 82\%, and 80\%, respectively. By increasing the intensity of the URLLC packets per slot, the eMBB reliability decreases to 71\% for a very high number of URLLC packets (70 packets per time slot) which, in fact, is less likely. The proposed algorithm surpasses the other solutions even for a high number of URLLC packets per slot, and the gap between our strategy and its closest competitor, PS, is significant. The proposed algorithm is 10\% more reliable than the PS case for 70 URLLC packets per time slot.

\section{Conclusion} \label{sec:con}
We investigated the coexistence problem of eMBB and URLLC in 5G-NR. We formulated the puncturing data rate problem for each eMBB user in order to study the impact of the incoming URLLC traffic, which must be scheduled immediately within mini-slots due to its extra low latency requirement. We proposed an optimization algorithm to enhance the minimum eMBB data rate per user and evaluated its performance with various loss functions, gNB transmission configuration regimes, and some state-of-the-art solutions. As a result, the proposed algorithm improves the data rate per eMBB user, even for the worst-case eMBB user. Besides, by applying the proposed optimization algorithm, the eMBB users experience a more reliable transmission than the other approaches. 
\bibliographystyle{IEEEtran}
\bibliography{refer}

\end{document}